\documentclass[12pt]{article}
\usepackage{latexsym}
\begin{document}
\rightline{IU-MSTP/74}
\rightline{August, 2007}

\vspace*{15mm}

\begin{center}
\LARGE{Operator ordering in Two-dimensional $N=1$\\
supersymmetry with curved manifold}

\vspace{12mm}
\def\thefootnote{\fnsymbol{footnote}}
\large{Nobuyuki MOTOYUI\footnote[1]{motoyui@mx.ibaraki.ac.jp} and
Mitsuru YAMADA\footnote[2]{m.yamada@mx.ibaraki.ac.jp}}
\def\thefootnote{\arabic{footnote}}

\vspace{12mm}
\large{Department of Mathematical Sciences, Faculty of Sciences,\\
Ibaraki University, Bunkyo 2-1-1, Mito, 310-8512, Japan}

\end{center}

\vspace{12mm}

\begin{abstract}
We investigate an operator ordering problem in two-dimensional $N=1$ supersymmetric model
which consists of $n$ real superfields. There arises an operator ordering problem when
the target space is curved. We have to fix the ordering in quantum operator properly to
obtain the correct supersymmetry algebra. We demonstrate that the super-Poincar\'{e}
algebra fixes the correct operator ordering. We obtain a supercurrent with correct operator
ordering and a central extension of supersymmetry algebra.
\end{abstract}


\newpage
\section{Introduction}

When we quantize classical field theory with curved target space, e.g. nonlinear sigma model,
there arises a problem of fixing of operator ordering in the various quantum observables.
We have to find out properly ordered quantum operator $\hat{F}$ from classical dynamical
variable $F(x,p)$. Operator ordering problem appears in various context.
For example, nonrelativistic Chern-Simons theory \cite{pp},
path integral in statistical mechanics \cite{egs} and so on.
There are several ways for this problem such as Weyl ordering \cite{das}
or self-adjoint extension of Hermitian operators \cite{rud}.

There also arise an operator ordering problem in quantizing a supersymmetric theory with
curved target space. The role of invariance under general coordinate transformation in
supersymmetric quantum mechanics in curved space was discussed before \cite{affr}.
In $N=2$ and $N=4$ supersymmetric quantum mechanics on a sphere, the energy spectrum
depends on the parameter which characterizes the operator ordering ambiguity
\cite{hllo-1,hllo-2}. In this paper, we consider a supersymmetrc field theory \cite{soh,wb}
in two dimensions. There also arises an operator ordering problem. Among several ways
for fixing of operator orders, we consider how symmetry decides the proper quantum operator.
In the previous paper \cite{my}, we have argued that the super-Poincar\'{e} algebra
gives a basis to fix the operator ordering properly in two-dimensional $N=2$ supersymmetry.
We can admit the supercharge operator $Q$ and $\bar{Q}$ have a correct operator order
when each component fields $\varphi$ satisfy the following relation:
\begin{eqnarray}
-i\lbrack \varphi, Q \rbrack_{\pm} -i\lbrack \varphi, \bar{Q} \rbrack_{\pm}
&=& \delta\varphi
  \label{eq:1-1}
\end{eqnarray}
in two-dimensional $N=2$ Wess-Zumino type model of $n$ chiral superfields $\phi^{i}$.
We will show that this is also true in two-dimensional $N=1$ supersymmetric theory.
We can admit the supercharge operator have a correct operator order when
each component fields satisfy the following relation:
\begin{eqnarray}
-i\lbrack \varphi, Q \rbrack_{\pm} &=& \delta\varphi
  \label{eq:1-2}
\end{eqnarray}
in two-dimensional $N=1$ supersymmetric model of $n$ real superfields $\phi^{i}$.

This paper is constructed as follows. In Sec. 2 we construct a two-dimensional $N=1$
supersymmetric model of $n$ real superfields with general nonflat target space.
In Sec. 3 we derive canonical quantization conditions through Dirac brackets. We fix
the operator order in $j^{\mu}$ by the relation (\ref{eq:1-2}) as the correct operator order.
Then we obtain the correct supersymmetry algebra.
Section 4 is devoted to conclusion of the work.
\section{Lagrangian and supercurrent}

We consider the two-dimensional $N=1$ supersymmetric theory.
Supercharge and covariant derivative operators are given by
\begin{eqnarray}
Q ~=~
i\left( \frac{\partial}{\partial\bar{\theta}} + i\gamma^{\mu}\theta\partial_{\mu}\right) &,&
\bar{Q} ~=~
-i\left( \frac{\partial}{\partial\theta} + i\bar{\theta}\gamma^{\mu}\partial_{\mu}\right),
  \label{eq:2-1}\\
D ~=~
\frac{\partial}{\partial\bar{\theta}} - i\gamma^{\mu}\theta\partial_{\mu} &,&
\bar{D} ~=~
-\left( \frac{\partial}{\partial\theta} - i\bar{\theta}\gamma^{\mu}\partial_{\mu}\right)
  \label{eq:2-2}
\end{eqnarray}
where $\theta$ is a two-dimensional Majorana spinor. These differential operators satisfy
the following anticommutators
\begin{eqnarray}
\{ Q,\bar{Q} \} &=& 2\gamma_{\mu}P^{\mu},
  \label{eq:2-3}\\
\{ D,\bar{D} \} &=& -2i\gamma^{\mu}\partial_{\mu}
  \label{eq:2-4}
\end{eqnarray}
and the other anticommutators vanish.
We use two-dimensional $\gamma$ matrices in the following representation:
\begin{eqnarray}
\gamma^{0}~=~\sigma_{2},~~
\gamma^{1}~=~-i\sigma_{1},~~
\gamma_{5}~=~\gamma^{0}\gamma^{1}~=~-\sigma_{3}.
  \label{eq:2-5}
\end{eqnarray}

A real superfield $\phi$ is given by
\begin{eqnarray}
\phi(x,\theta) &=& a(x) +\bar{\theta}\xi(x) +\frac{1}{2}\bar{\theta}\theta f(x),
  \label{eq:2-6}
\end{eqnarray}
where $a, \xi, f$ is a real scalar, a two-dimensional Majorana spinor and a auxiliary field
respectively. $\bar{\xi}$ represents $\xi^{T}\gamma^{0}$.

A supersymmetric Lagrangian consists of $n$ real superfields is given as
\begin{eqnarray}
{\cal L} &=&
\int d^{2}\theta \left\{ \frac{1}{2}g_{ij}(\phi)\bar{D}\phi^{i}D\phi^{j} + 2V(\phi) \right\},
  \label{eq:2-7}
\end{eqnarray}
where $g_{ij}$ is a metric of target space and $V$ is a superpotential. In terms of component fields
and eliminating the auxiliary fields, this Lagrangian becomes
\begin{eqnarray}
{\cal L} &=&
\frac{1}{2}g_{ij} \left( \partial_{\mu}a^{i}\partial^{\mu}a^{j}
+ \frac{i}{2}\bar{\xi}^{i}\gamma^{\mu}\stackrel{\leftrightarrow}{D_{\mu}}\xi^{j} \right) \nonumber\\
& &
+ \frac{1}{12}R_{ijkl} \bar{\xi}^{i}\xi^{k}\bar{\xi}^{j}\xi^{l}
- \frac{1}{2}\nabla_{i}V_{j}\bar{\xi}^{i}\xi^{j} - \frac{1}{2}g^{ij}V_{i}V_{j} \nonumber\\
&=&
\frac{1}{2}g_{ij} \left( \partial_{\mu}a^{i} \partial^{\mu}a^{j}
+ \frac{i}{2}\bar{\xi}^{i} \gamma^{\mu}\stackrel{\leftrightarrow}{\partial_{\mu}} \xi^{j} \right) \nonumber\\
& &
- \frac{i}{2}g_{ij,k}\bar{\xi}^{k}\gamma^{\mu}\xi^{i}\partial_{\mu}a^{j}
- \frac{1}{8} \left( g_{kl,mn} + g_{ij} {\Gamma^{i}}_{kl} {\Gamma^{j}}_{mn} \right) \bar{\xi}^{k}\xi^{l}\bar{\xi}^{m}\xi^{n} \nonumber\\
& &
- \frac{1}{2} \left( V_{ij} - {\Gamma^{k}}_{ij}V_{k} \right) \bar{\xi}^{i}\xi^{j}
- \frac{1}{2}g^{ij}V_{i}V_{j}.
  \label{eq:2-8}
\end{eqnarray}
The equation of motion for auxiliary field is
\begin{eqnarray}
f^{i} &=& \frac{1}{2} {\Gamma^{i}}_{jk}\bar{\xi}^{j}\xi^{k} - g^{ij}V_{j}.
  \label{eq:2-9}
\end{eqnarray}
We represent the derivatives by $a^{i}$ as follows
\begin{eqnarray}
V_{i} &=& \frac{\partial V(a)}{\partial a^{i}},
  \label{eq:2-10}\\
g_{ij,k} &=& \frac{\partial g_{ij}(a)}{\partial a^{k}}.
  \label{eq:2-11}
\end{eqnarray}
$R_{ijkl}$ is a Riemann tensor for the metric of target space.
$D_{\mu}\xi^{i}$ and $\nabla_{i}V_{j}$ means
\begin{eqnarray}
D_{\mu}\xi^{i} &=& \partial_{\mu}\xi^{i} + {\Gamma^{i}}_{mn}\xi^{m}\partial_{\mu}a^{n},
 \label{eq:2-12}\\
\nabla_{i}V_{j} &=& V_{ij} - {\Gamma^{m}}_{ij}V_{n}.
 \label{eq:2-13}
\end{eqnarray}

From this Lagrangian, canonical momentum operators conjugate to $a^{i}$ and $\xi^{i}$ are
\begin{eqnarray}
\pi_{a^{i}} &=&
\partial_{0}a^{j}g_{ji} - \frac{i}{2}g_{ij,k}\bar{\xi}^{k}\gamma^{0}\xi^{j},
 \label{eq:2-14}\\
\pi_{\xi^{i}} &=&
\frac{i}{2}g_{ij}\bar{\xi}^{j}\gamma^{0}.
 \label{eq:2-15}
\end{eqnarray}
Canonical energy-momentum tensor $T^{\mu\nu}$ is given as follows:
\begin{eqnarray}
T^{00} &=&
\frac{1}{2} \left( \partial_{0}a^{i}g_{ij}\partial_{0}a^{j} + \partial_{1}a^{i}g_{ij}\partial_{1}a^{j} \right)
- \frac{i}{4}g_{ij}\bar{\xi}^{i}\gamma^{1}\stackrel{\leftrightarrow}{\partial_{1}}\xi^{j} \nonumber\\
& &
+ \frac{i}{2}g_{ij,k}\bar{\xi}^{k}\gamma^{1}\xi^{j}\partial_{1}a^{i}
+ \frac{1}{8} \left( g_{kl,mn} + g_{ij} {\Gamma^{i}}_{kl} {\Gamma^{j}}_{mn} \right) \bar{\xi}^{k}\xi^{l}\bar{\xi}^{m}\xi^{n} \nonumber\\
& &
+ \frac{1}{2} \left( V_{ij} - {\Gamma^{k}}_{ij}V_{k} \right) \bar{\xi}^{i}\xi^{j}
+ \frac{1}{2}g^{ij}V_{i}V_{j},
 \label{eq:2-16}\\
T^{01} &=&
- \frac{1}{2} \left( \partial_{0}a^{i}g_{ij}\partial_{1}a^{j} + \partial_{1}a^{i}g_{ij}\partial_{0}a^{j} \right)
\nonumber\\
& &
+ \frac{i}{4} g_{ij,k} \partial_{1}a^{i}
\left( \bar{\xi}^{k}\gamma^{0}\xi^{j} - \bar{\xi}^{j}\gamma^{0}\xi^{k} \right)
- \frac{i}{4} g_{ij}\bar{\xi}^{j}\gamma^{0}\stackrel{\leftrightarrow}{\partial_{1}}\xi^{i}.
 \label{eq:2-17}
\end{eqnarray}
The supercurrent $J^{\mu}$ is  given by Noether procedure
\begin{eqnarray}
J^{\mu} &=& \bar{\eta}j^{\mu}, \nonumber\\
j^{\mu} &=&
\partial_{0}a^{i}g_{ij}\gamma^{0}\gamma^{\mu}\xi^{j}
+ \partial_{1}a^{i}g_{ij}\gamma^{1}\gamma^{\mu}\xi^{j}
+ iV_{i}\gamma^{\mu}\xi^{i}
 \label{eq:2-18}
\end{eqnarray}
where $\bar{\eta}$ is a parameter of supersymmetry transformation.
Particularly the time component of  $j^{\mu}$ becomes
\begin{eqnarray}
j^{0} &=&
\partial_{0}a^{i}g_{ij}\xi^{j}
- \partial_{1}a^{i}g_{ij}\gamma_{5}\xi^{j}
+ iV_{i}\gamma^{\mu}\xi^{i}.
 \label{eq:2-19}
\end{eqnarray}
The supercharge is defined by
\begin{eqnarray}
Q &=& \int_{-\infty}^{\infty} j^{0}(x) dx.
  \label{eq:2-20}
\end{eqnarray}
\section{Dirac bracket quantization and fixing of operator order}

Canonical momentum for $a$ and $\xi$ are given as (\ref{eq:2-14}), (\ref{eq:2-15}).
On canonical quantization, canonical momenta for $\xi$ give a primary
constraint:
\begin{eqnarray}
\chi_{\xi^{i}} &=& \pi_{\xi^{i}} - \frac{i}{2}g_{ij}\bar{\xi}^{j}\gamma^{0} ~=~ 0.
  \label{eq:3-1}
\end{eqnarray}
Poisson bracket for the constraint $\chi^{i}$ is
\begin{eqnarray}
\{ \chi^{i}, \chi^{j} \}_{P} &=& -ig_{ij}.
  \label{eq:3-2}
\end{eqnarray}
Therefore the constraint $\chi^{i}$ is the second class constraint.

Canonical quantization condition is given through the Dirac brackets.
There are seven nonzero Dirac brackets in 15 Dirac brackets. There are four independent
Dirac brackets in these seven nonzero Dirac brackets. They are given as follows:
\begin{eqnarray}
\{ a^{i},\pi_{a^{j}} \}_{D} &=& \delta^{i}_{j},
  \label{eq:3-3a}\\
\{ \xi^{i},\xi^{j} \}_{D} &=& -ig^{ij},
  \label{eq:3-3b}\\
\{ \xi^{i},\pi_{a^{j}} \}_{D} &=& -\frac{1}{2}g^{il}g_{lm,j}\xi^{m},
  \label{eq:3-3c}\\
\{ \pi_{a^{i}},\pi_{a^{j}} \}_{D} &=&
-\frac{i}{4}g^{kl}g_{km,i}g_{ln,j}\bar{\xi}^{m}\gamma^{0}\xi^{n}.
  \label{eq:3-3d}
\end{eqnarray}
Replacing these Dirac brackets with (anti) commutators divided by $i$,
we obtain the following canonical quantization conditions:
\begin{eqnarray}
\lbrack a^{i}(x,t), \partial_{0}a^{j}(y,t) \rbrack &=& g^{ij}(x,t) \cdot i\delta(x-y),
  \label{eq:3-4a}\\
\{ \xi^{i}(x,t), \xi^{j}(y,t) \} &=& -ig^{ij}(x,t) \cdot i\delta(x-y),
  \label{eq:3-4b}\\
\lbrack \xi^{i}(x,t), \partial_{0}a^{j}(y,t) \rbrack &=&
-\left( g^{jl}{\Gamma^{i}}_{lk}\xi^{k} \right) (x,t) \cdot i\delta(x-y),
  \label{eq:3-4c}\\
\lbrack \partial_{0}a^{i}(x,t), \partial_{0}a^{j}(y,t) \rbrack &=&
\left( \partial_{0}a^{m}g^{in}g^{kj}g_{mk,n} - g^{in}g_{nm,k}\partial_{0}a^{m}g^{kj} \right. \nonumber\\
& &
+ ig^{im}g^{jn}g_{mk,nl}\bar{\xi}^{l}\gamma^{0}\xi^{k} \nonumber\\
& &
\left. + ig^{im}g^{jn}g_{rs}{\Gamma^{r}}_{km}{\Gamma^{s}}_{ln}\bar{\xi}^{l}\gamma^{0}\xi^{k} \right)
(x,t)\cdot i\delta(x-y),
  \label{eq:3-4d}
\end{eqnarray}
and the other (anti) commutators are zero.

When we transfer from classical theory to quantum theory, we have to fix the order of operators.
We have to fix the order of the operators appear in $j^{\mu}$ and $T^{\mu\nu}$ to obtain
the correct supersymmetry algebra.

However there is no ordering problem in the case which the target space has a flat metric.
In this case, the supercurrent is
\begin{eqnarray}
j^{\mu} &=&
\partial_{0}a_{i}\gamma^{0}\gamma^{\mu}\xi^{i}
+ \partial_{1}a_{i}\gamma^{1}\gamma^{\mu}\xi^{i}
+ iV_{i}\gamma^{\mu}\xi^{i}.
  \label{eq:4-1a}
\end{eqnarray}
Canonical energy-momentum tensor is
\begin{eqnarray}
T^{00} &=&
\frac{1}{2} \left( \partial_{0}a_{i}\partial_{0}a^{i} + \partial_{1}a_{i}\partial_{1}a^{i} \right)
- \frac{i}{4}\bar{\xi}_{i}\gamma^{1}\stackrel{\leftrightarrow}{\partial_{1}}\xi^{i}
+ \frac{1}{2}V_{ij}\bar{\xi}^{i}\xi^{j} + \frac{1}{2}V_{i}V^{i},
  \label{eq:4-1b}\\
T^{01} &=&
- \frac{1}{2} \left( \partial_{0}a_{i}\partial_{1}a^{i} + \partial_{1}a_{i}\partial_{0}a^{i} \right)
- \frac{i}{4}\bar{\xi}_{i}\gamma^{0}\stackrel{\leftrightarrow}{\partial_{1}}\xi^{i}.
  \label{eq:4-1c}
\end{eqnarray}
In this case we can obtain the correct supersymmetry algebra regardless of the order of
the operators because canonical quantization conditions become as usual:
\begin{eqnarray}
\lbrack a^{i}(x,t), \partial_{0}a^{j}(y,t) \rbrack &=& \eta^{ij}(x,t) \cdot i\delta(x-y),
  \label{eq:4-1d}\\
\{ \xi^{i}(x,t), \xi^{j}(y,t) \} &=& -i\eta^{ij}(x,t) \cdot i\delta(x-y),
  \label{eq:4-1e}
\end{eqnarray}
and the other (anti) commutators are zero.
Therefore there is no ordering problem in the supercurrent $j^{\mu}$ when
the target space has a flat metric.

On the other hand, there becomes a ordering problem when the target space has a nonflat metric.
In this case, we have to fix the orders of the operators appear in $j^{\mu}$ correctly
to obtain the correct supersymmetry algebra. We rely upon the supersymmetry to fix the orders of
the operators: when the theory has a supersymmetry, it gives the correct order of the operators.
To fix the operator orders correctly, we require that each component fields $\varphi$ satisfy
the following relation:
\begin{eqnarray}
-i\lbrack \varphi, Q \rbrack_{\pm} &=& \delta\varphi.
  \label{eq:4-2}
\end{eqnarray}
The supersymmetry transformations are given as follows
\begin{eqnarray}
\delta a^{i} &=& \bar{\eta}\xi^{i},
  \label{eq:4-3a}\\
\delta \xi^{i} &=&
\left( \frac{1}{2} {\Gamma^{i}}_{jk}\bar{\xi}^{j}\xi^{k} - g^{ij}V_{j}
- i\partial_{\mu}a^{i}\gamma^{\mu} \right) \eta
  \label{eq:4-3b}
\end{eqnarray}
where $\eta$ is a parameter of the supersymmetry transformation.
Among several operator orders, we take the operator order which satisfies
the above relations as the correct operator order of supercurrent $j^{\mu}$. The above
representation of supercurrent (\ref{eq:2-19}) does not satisfy the relation (\ref{eq:4-2}).
The following representation of supercurrent also does not satisfy the relation (\ref{eq:4-2}):
\begin{eqnarray}
j^{0} &=&
\xi^{j}g_{ij}\partial_{0}a^{i}
- \partial_{1}a^{i}g_{ij}\gamma_{5}\xi^{j}
+ iV_{i}\gamma^{\mu}\xi^{i}.
 \label{eq:4-3c}
\end{eqnarray}
We have to symmetrize the terms which involve $\partial_{0}a^{i}$ to satisfy the relation
(\ref{eq:4-2}):
\begin{eqnarray}
j^{0} &=&
\frac{1}{2} \left( \partial_{0}a^{i}g_{ij}\xi^{j} + \xi^{j}g_{ij}\partial_{0}a^{i} \right)
- \partial_{1}a^{i}g_{ij}\gamma_{5}\xi^{j}
+ iV_{i}\gamma^{\mu}\xi^{i}.
 \label{eq:4-3d}
\end{eqnarray}
This representation of supercurrent satisfy the relation (\ref{eq:4-2}).
Therefore this supercurrent gives the correct supersymmetry algebra:
\begin{eqnarray}
\{ Q,\bar{Q} \} &=& 2\gamma^{\mu}P_{\mu} + \gamma_{5}T.
  \label{eq:4-4}
\end{eqnarray}
The representations of canonical energy-momentum tensor appear in $P_{\mu}$ are given as
(\ref{eq:2-16}) and (\ref{eq:2-17}).
The central charge $T$ is given as the difference of superpotential:
\begin{eqnarray}
T &=&
2i\int_{-\infty}^{\infty} \partial_{1}\left( V(a^{i}(x)) \right)dx \nonumber\\
&=&
2i\left\{ V(a^{i}(x=\infty)) - V(a^{i}(x=-\infty)) \right\} \nonumber\\
&\equiv&
2i\Delta V.
  \label{eq:4-5}
\end{eqnarray}
\section{Conclusion}

When we quantize classical field theory with curved target space, there arises a problem of fixing of operator ordering in the various quantum observables. This is also true in supersymmetric theory.
We have to fix the ordering in quantum operator properly to obtain the correct supersymmetry
algebra. We investigated a operator ordering problem in two-dimensional $N=1$ supersymmetric
model which consists of $n$ real superfields.
In the previous paper \cite{my}, we have argued that the supersymmetry
gives a basis to fix the operator ordering properly in two-dimensional $N=2$ supersymmetry.
We can admit that the super-Poincar\'{e} algebra gives the correct operator ordering
in two-dimensional $N=2$ supersymmetry. Here we showed that it is also true in two-dimensional
$N=1$ supersymmetry.  We can take the supercurrent operator (\ref{eq:4-3d}) which was
symmetrized in the terms that involve $\partial_{0}a^{i}$ as a proper quantum operator.
It gives the correct supersymmetry algebra with a central charge. It may be also applied to
higher dimensional case. The proper operator ordering may be determined by supersymmetry:
the super-Poincar\'{e} algebra may gives the correct operator ordering.

\end{document}